\documentclass[reprint,NumberedRefs]{JASA}

\usepackage{siunitx}

\begin{document}

\title{\textit{In situ} estimation of the acoustic surface impedance using simulation-based inference}

\author{Jonas M. Schmid}
\email{jonas.m.schmid@tum.de}
\author{Johannes D. Schmid}
\author{Martin Eser}
\author{Steffen Marburg}
\affiliation{Chair of Vibroacoustics of Vehicles and Machines, Technical University of Munich, Garching near Munich, 85748, Germany}

\date{\today}

\begin{abstract}
Accurate acoustic simulations of enclosed spaces require precise boundary conditions, typically expressed through surface impedances for wave-based methods. Conventional measurement techniques often rely on simplifying assumptions about the sound field and mounting conditions, limiting their validity for real-world scenarios. To overcome these limitations, this study introduces a Bayesian framework for the \textit{in situ} estimation of frequency-dependent acoustic surface impedances from sparse interior sound pressure measurements. The approach employs simulation-based inference, which leverages the expressiveness of modern neural network architectures to directly map simulated data to posterior distributions of model parameters, bypassing conventional sampling-based Bayesian approaches and offering advantages for high-dimensional inference problems. Impedance behavior is modeled using a damped oscillator model extended with a fractional calculus term. The framework is verified on a finite element model of a cuboid room and further tested with impedance tube measurements used as reference, achieving robust and accurate estimation of all six individual impedances. Application to a numerical car cabin model further demonstrates reliable uncertainty quantification and high predictive accuracy even for complex-shaped geometries. Posterior predictive checks and coverage diagnostics confirm well-calibrated inference, highlighting the method’s potential for generalizable, efficient, and physically consistent characterization of acoustic boundary conditions in real-world interior environments.
\end{abstract}


\maketitle



\section{\label{sec:Introduction} Introduction}
With advances in computational power and numerical algorithms, wave-based simulation techniques such as the finite element method (FEM) or the boundary element method (BEM) have become essential tools for predicting sound fields in enclosed spaces~\cite{Prinn.2023}. These methods support acoustic design already in the early development stages and are applied in a wide range of contexts, including music rehearsal and performance spaces, recording studios, lecture halls, office environments, as well as automotive and aircraft cabins~\cite{Gurbuz.2023}. As emphasized by Aretz and Vorländer~\cite{Aretz.2014b}, the accuracy and robustness of such simulations critically depend on the precise characterization of all acoustically interacting surfaces~\cite{Vorlander.2013, Fratoni.2025}, which is a major source of uncertainties~\cite{Thydal.2021}. In wave-based models, boundary conditions are commonly described by the complex-valued surface impedance, or equivalently by its reciprocal, the surface admittance~\cite{Marburg.2008}. These quantities inherently account for both amplitude and phase information, both being indispensable for physically accurate predictions~\cite{Marburg.1999}. 
Traditionally, energy-based quantities such as the absorption coefficient have been employed in room acoustics to describe the fraction of acoustic energy absorbed by a boundary. While extensive datasets of random-incidence absorption coefficients are available in literature, these data provide only magnitude, but lack the corresponding phase information. This limitation is particularly critical in small spaces and the low frequency range, where room modes dominate~\cite{CardosoSoares.2022}. Despite the increasing relevance of wave-based methods in room acoustics, comprehensive datasets of the surface impedance remain scarce~\cite{Fratoni.2025}. Three different methods have emerged to measure the absorptive properties of a material.
\paragraph*{Reverberation room method:}
The reverberation room method, standardized in ISO 354~\cite{ISO354.2003}, determines the acoustic absorption coefficient of materials by measuring the change in reverberation time of a diffuse sound field when a test specimen is introduced into a reverberation chamber. This method is widely used for characterizing materials under diffuse-field conditions. However, reverberation chambers are not always available in laboratory settings, and the diffuse-field assumption becomes invalid at low frequencies, leading to significant uncertainties~\cite{Nolan.2018}.

\paragraph*{Impedance tube method:}
The impedance tube method, standardized in ISO~10534-2~\cite{ISO105342.1998}, is a well-established method for characterizing the normal-incidence surface impedance. It relies on evaluating the transfer function between two microphones positioned inside a rigid tube, with the test specimen mounted at its termination. Despite its practical utility, the method is inherently restricted to normal incidence of plane waves, which contrasts with real-world acoustic conditions where surfaces are exposed to complex sound fields and waves impinge from multiple directions~\cite{Eser.2025}. Moreover, the use of small specimen sizes and idealized mounting conditions in the tube often fail to reflect realistic installation scenarios. As demonstrated by Brandão et al.~\cite{Brandao.2013}, materials that appear to perform adequately in impedance tube measurements may display markedly different acoustic behavior under practical operating conditions.

\paragraph*{\textit{In situ} methods:}
The limitations of the two aforementioned methods highlight the need for \textit{in situ} methods that allow for non-destructive characterization of materials under realistic mounting conditions, without relying on laboratory setups or assumptions about wave incidence~\cite{Brandao.2015}. To this end, several approaches have been developed, most of which are formulated as inverse problems. In these methods, sound pressure is measured in the vicinity of the sample, and the surface impedance or absorption coefficient is subsequently estimated by solving an inverse problem based on physical models of wave propagation~\cite{Tamura.1990}. 
The sound field can be decomposed into elementary wave functions including plane waves~\cite{Nolan.2020} and spherical waves~\cite{Richard.2017}. 
Microphone arrays enable simultaneous sound pressure measurements at multiple positions and are commonly implemented in different configurations such as double-layered planar arrays~\cite{Hald.2019} or spherical arrays~\cite{Richard.2019b}. These measurements are typically performed in (semi-)anechoic chambers under free-field conditions to ensure well-defined incidence and to minimize undesired wall reflections~\cite{Eser.2023, Brandao.2011}. More recently, neural network–based approaches have been introduced to compensate for edge diffraction effects caused by the finite size of test samples~\cite{Emmerich.2025, MullerGiebeler.2024, Zea.2023}. Alternatively, pressure–particle velocity probes can be employed to directly capture sound pressure and particle velocity in close proximity to the sample surface, with the measured data subsequently combined with the aforementioned techniques to estimate boundary properties~\cite{BrieredelaHosserayeBaltazar.2022, MullerTrapet.2013}. Generally, the accuracy of methods based on wave propagation models strongly depends on the complexity of the selected model and inherent simplifying assumptions, with these limitations becoming particularly pronounced at low frequencies and in geometrically complex environments~\cite{Luo.2020}.

To overcome these limitations, discretization-based methods have been employed in which numerical methods such as the FEM or BEM serve as forward models to compute the sound pressure field in an acoustic domain for given boundary conditions. Such approaches are particularly advantageous for complex geometries and in cases where the domain is well known or a detailed geometric model already exists, for example in digital twin applications or model updating. Dutilleux et al.~\cite{Dutilleux.2002} have combined a global evolutionary optimization strategy with the finite difference method and FEM to estimate boundary impedances of rectangular acoustic domains in the low-frequency range. Anderssohn and Marburg~\cite{Anderssohn.2007} have employed a FEM model together with a two-step nonlinear optimization scheme to recover spatially non-uniform boundary admittances from interior sound pressure measurements. Similarly, Luo et al.~\cite{Luo.2020} have used BEM in conjunction with iterative optimization procedures. 
Prinn et al.~\cite{Prinn.2021} have introduced an eigenvalue approximation method combined with optimization to estimate normal incidence surface impedances at modal frequencies for the case of spatially uniform surfaces. 
More recently, neural network–based approaches have gained attention for estimating acoustic boundary conditions and predicting sound radiation~\cite{Schmid.2025}. Schmid et al.~\cite{Schmid.2024} and Xia et al.~\cite{Xia.2024} have proposed physics-informed neural networks to infer acoustic boundary properties directly from pressure measurements, demonstrating promising accuracy. Nevertheless, these approaches remain limited to simple geometries.

Inferring surface impedances from acoustic pressure measurements constitutes an ill-posed inverse problem that is highly sensitive to noise, measurement uncertainty, and modeling errors. These uncertainties may originate from various sources, including inaccuracies in estimating the positions of sound sources and sensors~\cite{Brandao.2013}, spurious reflections between the measurement setup and the sample~\cite{MullerTrapet.2013}, and inaccuracies in calibration procedures~\cite{Brandao.2011}. Additional errors may result from the non-linear and frequency-dependent behavior of sensors and sound sources~\cite{Vorlander.2013}. Moreover, model-related uncertainties arise from simplifying assumptions, such as idealizing sound sources~\cite{Thydal.2021} or oversimplified representations of the surface reflection characteristics~\cite{Dragonetti.2017}.
Deterministic approaches typically address the ill-posedness of the inverse problem by incorporating regularization techniques, such as Tikhonov regularization~\cite{Nolan.2020} or compressive sensing~\cite{Shen.2024}, which essentially reformulate the problem to promote stability and ensure a unique solution. However, the choice of regularization parameters is problem-dependent and can strongly bias the solution~\cite{Hansen.1992}. Furthermore, deterministic methods generally yield single point estimates and do not provide a quantification of the associated uncertainty, which limits their reliability and robustness, particularly when incorporated measurement data are sparse and noisy. 

\subsection{Bayesian inference in acoustics}
To overcome these limitations, a Bayesian framework is employed in which all quantities are represented as random variables. The associated randomness is used to reflect the degree of belief regarding their possible values. In contrast to point estimates, Bayesian inference provides full probability distributions of unknown model parameters, thereby offering a comprehensive characterization of the parameter space. This enables principled representation of uncertainty, may reveal dependencies between parameters, and ensures robustness even when measurement data are sparse or contaminated by noise~\cite{Xiang.2020}.

In recent years, Bayesian inference has been increasingly applied to a broad range of problems in acoustics. Applications include the characterization of porous materials~\cite{Fackler.2018}, fibrous materials~\cite{Yang.2025}, poroelastic materials~\cite{Cuenca.2022}, sound field reconstruction~\cite{Schmid.2021}, estimation of sound energy decay~\cite{Balint.2019}, sound absorber~\cite{Xiang.2022} and acoustic coating design~\cite{Modur.2025}.
In the context of surface impedance characterization, Bockmann et al.~\cite{Bockman.2015} have applied a Bayesian optimization algorithm to estimate the frequency-dependent surface impedance of a sample in a duct. 
More recently, Chen et al.~\cite{Chen.2022} have utilized Bayesian inference to jointly estimate the model order and parameters of a multipole admittance model for boundary admittance determination in an impedance tube setup. 
\begin{figure*}[htb]
	\begin{center}
		\includegraphics[width = \textwidth]{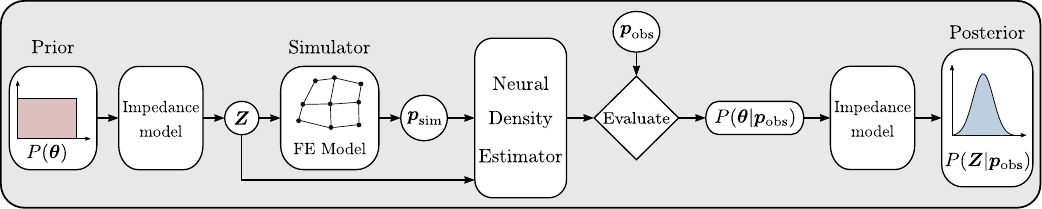}
		\caption{\label{fig:SBI_scheme}{Schematic overview of the proposed SBI framework for acoustic surface impedance estimation $\boldsymbol{Z}$. Parameters $\boldsymbol{\theta}$ are sampled from a prior distribution and mapped through an impedance model into impedance boundary conditions for a FEM simulator, yielding simulated pressure fields. A neural density estimator is trained on these simulations to approximate the posterior distribution of the parameters given observed measurements $\boldsymbol{p}_{\text{obs}}$. The resulting posterior is then propagated through the impedance model to obtain the final impedance estimates with quantified uncertainty.}}
	\end{center}
\end{figure*} 
All previously discussed Bayesian approaches for surface impedance characterization remain limited to impedance tube configurations, where sound propagation is restricted to plane waves. In contrast, Eser et al.~\cite{Eser.2023} have introduced a Bayesian inference framework based on sequential frequency transfer that enables efficient \textit{in situ} estimation of the frequency-dependent surface admittance with quantified uncertainties. However, the forward model employed in their study is inherently restricted to free-field conditions. Schmid et al.~\cite{Schmid.2023} have proposed a Bayesian framework for full \textit{in situ} estimation of piece-wise constant boundary admittances, providing complete posterior distributions. By combining a FEM-based forward model with gradient-based Bayesian sampling, the approach has successfully been demonstrated on a 2D car cabin model with complex geometry. Building on this direction, Wulbusch et al.~\cite{Wulbusch.2024} have presented a related method that achieves high accuracy in room acoustics applications when the data are consistent with the model, but its performance deteriorates in frequency ranges where discrepancies between model and data occur. 

However, these approaches perform inference at each frequency independently, without imposing a physically meaningful frequency-dependent impedance model across the frequency spectrum. Impedance models typically involve at least four parameters per local impedance surface~\cite{Mondet.2020}, which quickly scales to a large number of unknown parameters and thus a high-dimensional inference problem. Conventional Bayesian sampling strategies, such as Markov-chain Monte Carlo (MCMC) sampling~\cite{Metropolis.1953}, are known to struggle to adequately explore high-dimensional posterior spaces, often resulting in poor mixing and inefficient convergence due to the low acceptance rate of samples. In general, sampling-based methods are computationally demanding, as every posterior sample requires a full evaluation of the forward model. This effort becomes prohibitive when high-fidelity numerical forward models are employed. More advanced gradient-based sampling strategies increase this burden even further, since additional gradient information needs to be computed~\cite{Schmid.2023}.

To overcome these limitations, the proposed framework employs simulation-based inference (SBI)~\cite{Cranmer.2020}. SBI originates from approximate Bayesian computation approaches but employs neural networks to learn a direct mapping from simulated data to posterior distributions. Recent advances in neural network technology and the expressive power of modern neural network architectures enable to capture complex posterior distributions, including high-dimensional, multi-modal, and non-Gaussian structures, which are often inaccessible to traditional sampling-based approaches. 
Up to now, SBI has successfully been applied to inverse problems in diverse scientific disciplines such as particle physics~\cite{Brehmer.2021}, astrophysics~\cite{MishraSharma.2022} and neuroscience~\cite{Boelts.2023}. Motivated by the advantages of SBI for handling high-dimensional and computationally demanding inference tasks, this work aims to transfer SBI in to its first application in the field of acoustics. Specifically, a SBI framework for the \textit{in situ} estimation of frequency-dependent, piece-wise constant surface impedances is introduced.

This paper starts with a concise methodological overview of SBI, followed by a description of the underlying problem of \textit{in situ} material characterization and the frequency-dependent impedance model in Sec.~\ref{sec:Problem statement}. The proposed framework is first evaluated on a benchmark cuboid room problem in Sec.~\ref{sec: Cuboid room}, before being applied to a complex car cabin model in Sec.~\ref{sec:Car cabin problem}. Finally, Sec.~\ref{sec:Conclusions} summarizes the main findings and discusses the advantages and limitations of the approach.

\section{\label{sec:Simulation-based Inference} Simulation-based Inference}
Simulation-based inference applies Bayesian parameter estimation to problems where the forward model is defined through a simulator. The foundation of Bayesian inference is Bayes’ theorem, which in the context of parameter estimation is expressed as
\begin{align}
	P(\boldsymbol{\theta} \mid \boldsymbol{p}_{\text{obs}}) \propto P(\boldsymbol{p}_{\text{obs}} \mid \boldsymbol{\theta}) \, P(\boldsymbol{\theta}),
\end{align}
where the prior probability $P(\boldsymbol{\theta})$ encodes physically meaningful knowledge about the model parameters $\boldsymbol{\theta}$, informed by expert knowledge or literature, before any data are observed. The likelihood $P(\boldsymbol{p}_{\text{obs}} \mid \boldsymbol{\theta})$ quantifies the probability of observing measurement data $\boldsymbol{p}_{\text{obs}}$ given a particular parameter set $\boldsymbol{\theta}$. Their product yields the posterior probability $P(\boldsymbol{\theta} \mid \boldsymbol{p}_{\text{obs}})$, which represents the updated belief about the parameters $\boldsymbol{\theta}$ after incorporating both prior knowledge and observed data $\boldsymbol{p}_{\text{obs}}$.

In conventional Bayesian analysis, the likelihood is typically assigned following the principle of maximum entropy~\cite{Xiang.2020}, ensuring that no additional information beyond what is already known is introduced. For complex simulation-based forward models, it is generally not possible to derive a closed-form expression for the likelihood without imposing simplifying assumptions, which introduces bias~\cite{Cranmer.2020}. A key advantage of simulation-based approaches is that, while the likelihood cannot be written down explicitly, it is straightforward to sample from it by running the simulator. For a given parameter set $\boldsymbol{\theta}$, a single simulation run produces a sample $\boldsymbol{p}_{\text{sim}} \sim P(\boldsymbol{p}_{\text{sim}} \mid \boldsymbol{\theta})$. Repeating this process for $N_{\text{sim}}$ parameter configurations yields a dataset $\left\{ \bigl( \boldsymbol{\theta}_j, \boldsymbol{p}_{\text{sim},j} \bigr) \right\}_{j=1,\ldots,N_{\text{sim}}}$. SBI exploits this dataset to train a neural network that learns the probabilistic relationship between simulated data and corresponding parameters. The target posterior distribution $P(\boldsymbol{\theta} \mid \boldsymbol{p}_{\text{obs}})$ for a specific observation $\boldsymbol{p}_{\text{obs}}$ can then be interpreted as the conditional distribution of $\boldsymbol{\theta}$ when fixing the data to $\boldsymbol{p}_{\text{obs}}$.
In this framework, neural posterior estimation~\cite{Papamakarios.2016} is employed, which directly approximates the posterior probability of $\boldsymbol{\theta}$ conditioned on simulated data $\boldsymbol{p}_{\text{sim}}$. The neural density estimator, denoted as $q(\boldsymbol{\theta}\mid\boldsymbol{p}_{\text{sim}})~\approx~P(\boldsymbol{\theta}\mid\boldsymbol{p}_{\text{sim}})$, is trained by minimizing the negative log-posterior probability
\begin{align}
	\mathcal{L} = -\frac{1}{N_{\text{sim}}} \sum_{j=1}^{N_{\text{sim}}} \log q(\boldsymbol{\theta}_{j} \mid \boldsymbol{p}_{\text{sim}, j})
\end{align}
as loss function, which encourages the network to assign high probability to the parameters in the training dataset that have generated the corresponding simulated observations. The underlying architecture leverages flexible conditional generative models, in particular normalizing flows, enabling the approximation of complex posterior distributions~\cite{Papamakarios.2021}. The central concept of normalizing flows is to construct a complex target posterior distribution by successively transforming a simple base distribution, typically a Gaussian distribution, through a sequence of invertible mappings. Once trained, this framework enables efficient sampling from the approximate posterior by first drawing samples from the base distribution and then propagating them through the learned transformations. A key advantage of neural posterior estimation is its fully amortized inference capability. Once the network has been trained, it can efficiently generate samples from the posterior distribution for any new observation $\boldsymbol{p}_{\text{obs}}$ without the need for further training, additional simulations, or computationally expensive iterative procedures such as MCMC sampling or variational inference~\cite{Deistler.18.08.2025}.

A schematic overview of the proposed SBI framework for acoustic surface impedance estimation is shown in Fig.~\ref{fig:SBI_scheme}. Samples of the parameter vector $\boldsymbol{\theta}$ are drawn from a prior distribution and transformed via an impedance model into surface impedance boundary conditions for a FEM simulator, which outputs simulated pressure fields $\boldsymbol{p}_{\text{sim}}$. These simulations are used to train the inference network that approximates the posterior distribution of the parameters conditioned on the observed measurements. Finally, the inferred posterior is mapped back through the impedance model to provide impedance estimates together with quantified uncertainty.

\subsection{\label{subsec:Posterior diagnostics} Posterior diagnostics}
Once a posterior distribution has been obtained, it is crucial to evaluate the quality of the trained posterior estimator~\cite{Talts.18.04.2018}. A variety of diagnostic techniques are available, each targeting specific aspects of potential misspecification or miscalibration.

\paragraph*{Posterior predictive check:}
A posterior predictive check (PPC) provides a qualitative assessment of the adequacy of the posterior estimator by comparing simulated predictions with the observed measurements. $N_{\text{ppc}}$ parameter samples are drawn from the estimated posterior distribution and passed through the simulator to generate posterior predictive realizations. A well-calibrated posterior is expected to produce predictive samples that closely match the observed data, thereby demonstrating consistency between the inferred parameters, the simulator, and the measurements.

\paragraph*{Coverage diagnostic methods:}
Beyond predictive checks, a more rigorous evaluation of posterior quality can be achieved through coverage diagnostics, which assess the statistical calibration of the estimated posterior to ensure that uncertainty estimates are neither overconfident nor underconfident. In this work, calibration is examined locally for individual observations using the state-of-the-art local classifier two-sample test (L-C2ST) method~\cite{Linhart.2023}. To construct the calibration dataset, $N_{\text{cal}}$ samples are drawn from the prior, and the simulator is evaluated for each sample, yielding pairs of ground-truth parameters and simulated data. Posterior inference is then performed for each calibration sample in this set, resulting in $N_{\text{cal}}$ estimated posteriors. The key idea of L-C2ST is to train a binary classifier to distinguish between samples from the true posterior and the estimated posterior. If the classifier can reliably separate the two, this indicates a mismatch and thus flawed inference, whereas classification probabilities close to chance level (i.e. 0.5) suggest that the estimated posterior is well calibrated.

\section{\label{sec:Problem statement} Problem statement}
The acoustic pressure field $p(\boldsymbol{x})$ in the computational domain $\Omega$ with spatial dimension $d$ and spatial coordinate $\boldsymbol{x}$, subject to a source term $Q(\boldsymbol{x})$ is governed in frequency domain by the inhomogeneous Helmholtz equation, 
\begin{align}
	\nabla^2 p(\boldsymbol{x}) + k^2 p(\boldsymbol{x}) = -Q(\boldsymbol{x}), \qquad \boldsymbol{x} \in \Omega \subset \mathbb{R}^{d},
	\label{eq:Helmholtz}
\end{align}
where $\nabla^2$ denotes the Laplace operator and $k = \omega / c$ is the acoustic wavenumber, defined by the angular frequency $\omega = 2\pi f$ and the speed of sound $c$ in air. Eq.~\eqref{eq:Helmholtz} is derived under the assumptions of harmonic time dependence $e^{-\mathrm{i}\omega t}$ (with $\mathrm{i}^2=-1$), small perturbations of the acoustic field, and an inviscid fluid without mean flow. In this work, Robin boundary conditions are employed, relating the acoustic pressure to its normal derivative, or equivalently to the normal particle velocity $v_n$. They are most conveniently expressed through the acoustic surface impedance
\begin{align}
Z(\boldsymbol{x}) = \frac{p(\boldsymbol{x})}{v_n(\boldsymbol{x})}, \qquad \boldsymbol{x} \in \Gamma \subset \mathbb{R}^{d-1},
	\label{eq:Robin}
\end{align}
where $\Gamma$ denotes the boundary surface. For consistency, all impedances are normalized with respect to the characteristic impedance of air, $Z_0 = \rho_0 c$, with $\rho_0$ denoting the ambient fluid density. The real part of $Z$, referred to as the resistance, characterizes dissipative energy losses, while the imaginary part, known as the reactance, quantifies effects arising from temporary energy storage~\cite{Marburg.2011}. In general, surface impedance values may vary with incidence angle in complex sound fields. The widely used assumption of a locally reacting boundary implies that sound propagation inside the boundary material is restricted to the surface-normal direction, while tangential propagation is neglected. Together with the condition of a monotonically increasing reactance, these serve as the basic assumptions of the impedance model~\cite{Mondet.2020} introduced subsequently.

\subsection{\label{subsec:Impedance model} Impedance model}
A generalized frequency-dependent impedance model is employed, extending the classical damped oscillator formulation with an additional fractional calculus term to capture complex frequency-dependent behavior. Mondet et al.~\cite{Mondet.2020} have introduced this model and have demonstrated its flexibility in representing a wide range of common absorbers, including hard and soft porous materials, membrane absorbers, as well as perforated and microperforated panels. By means of constrained optimization, the model has successfully been applied to retrieve complex-valued surface impedances from statistical absorption coefficients~\cite{Mondet.2020}. A key advantage of this formulation is that it inherently guarantees physical consistency, as it respects passivity ($\Re(Z)\geq 0$), causality, and real-valuedness in the time domain. The impedance model can be expressed as~\cite{Mondet.2020}
\begin{align}
	Z(\omega) =  R + M \cdot (\mathrm{i}\omega) + K \cdot (\mathrm{i}\omega)^{-1} + G \cdot (\mathrm{i}\omega)^{\gamma}, 
\end{align}
with $R, M, K, G \geq 0$ and $-1 \leq \gamma \leq 1$. It consists of a frequency-independent resistive term $R$, a mass-related term $M \cdot (\mathrm{i}\omega)$, a compliance term $K \cdot (\mathrm{i}\omega)^{-1}$, and a fractional calculus term $G \cdot (\mathrm{i}\omega)^{\gamma}$, the latter accounting for frequency-dependent resistance and viscoelastic dissipation effects. In the low-frequency regime below $\SI{500}{Hz}$, the stiffness and resistive contributions significantly dominate over the inertial term~\cite{Mondet.2020}, allowing the mass component to be omitted without compromising accuracy. The model therefore simplifies to
\begin{align}
	Z(\omega) = R + K \cdot (\mathrm{i}\omega)^{-1} + G \cdot (\mathrm{i}\omega)^{\gamma}.
	\label{eq:Impedance_model}
\end{align}

\section{\label{sec: Cuboid room}Cuboid room problem}
\begin{figure}[htb]
	\centering
	\includegraphics[width=0.93\columnwidth]{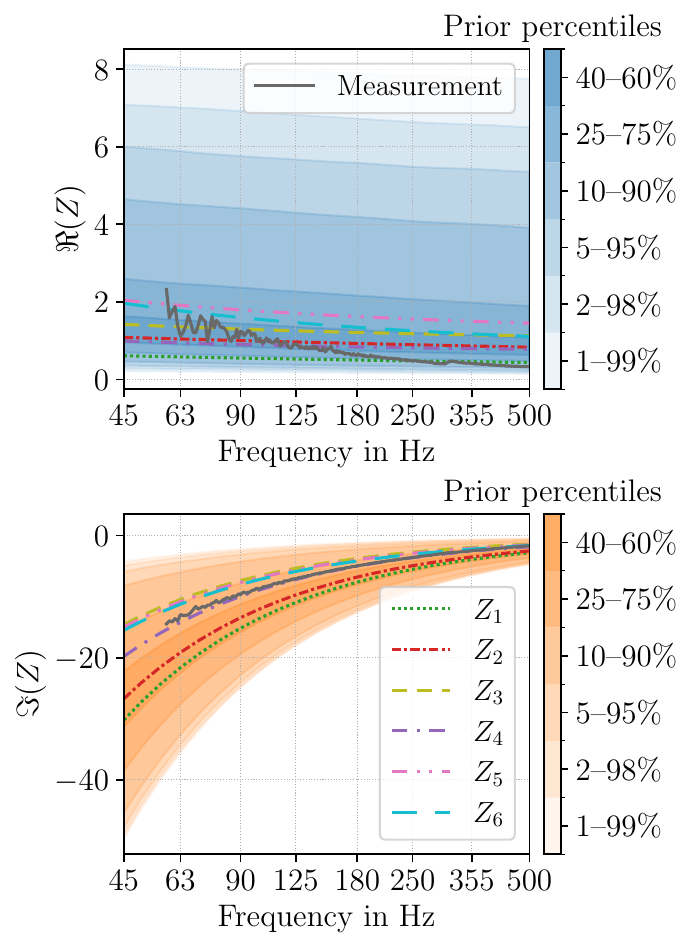}
	\caption{Real (top) and imaginary (bottom) parts of the reference surface impedances $Z_1$--$Z_6$ across the considered frequency range. The shaded regions represent the percentiles of the assigned prior distributions.}
	\label{fig:Priors+True_Z}
\end{figure}

In this study, a benchmark model of a three-dimensional cuboid room is investigated with dimensions $L_x=\SI{0.963}{m}$, $L_y=\SI{0.975}{m}$, and $L_z=\SI{2.075}{m}$, corresponding to a total volume of $\SI{1.95}{m^3}$.
\begin{figure}[tbh]
	\centering
	\includegraphics[width=0.9\columnwidth]{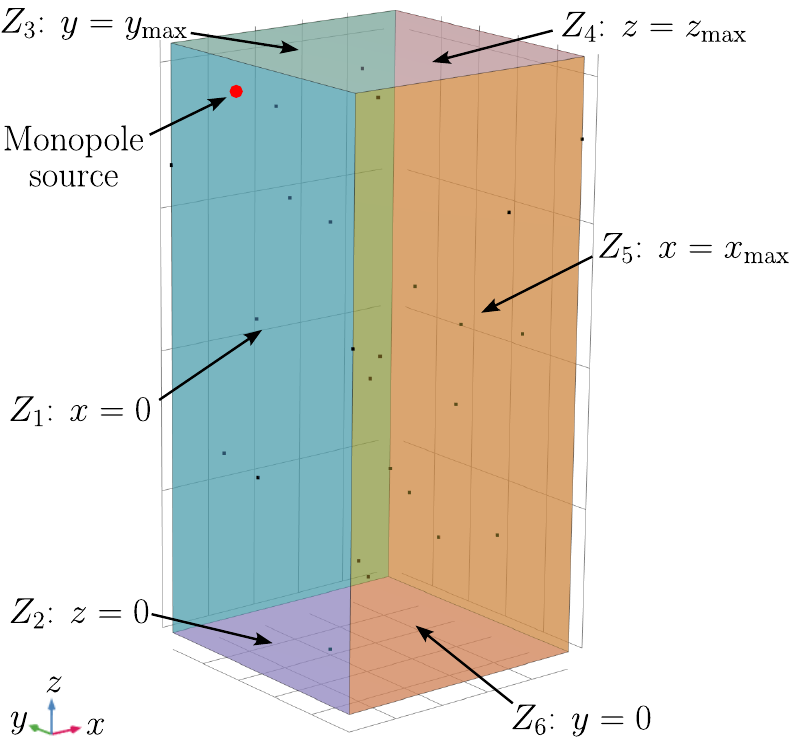}
	\caption{Geometric model of the cuboid room under investigation. Boundary surfaces with different impedance conditions are distinguished by color. The monopole source position is marked in red, while the black dots indicate the selected observation positions.}
	\label{fig:Model_Cuboid_room}
\end{figure}
The chosen geometry is representative of compact acoustic enclosures, such as soundproof phone cabins commonly found in office environments or recording and practice booths used in music studios. The domain is filled with air, modeled with a density of $\rho_0=\SI{1.2}{\kg\per\cubic\m}$ and a speed of sound of $c=\SI{343}{\m \per \s}$. Since the analysis is restricted to the low-frequency regime, the acoustic source can be represented by a monopole with a prescribed volume flow rate of $Q=\SI{1.0e-4}{\cubic\m\per\s}$. The point source is positioned at $\boldsymbol{x} = (\SI{0.15}{m},\, \SI{0.825}{m},\, \SI{1.925}{m})$. Fig.~\ref{fig:Model_Cuboid_room} depicts the geometric model of the enclosure. A distinct local surface impedance $Z_1 ,Z_2 ... Z_6$ is assigned to each piece-wise constant boundary surface, represented by different colors in Fig.~\ref{fig:Model_Cuboid_room}. To generate synthetic measurement data, frequency-dependent surface impedances are computed using the impedance model defined in Eq.~\eqref{eq:Impedance_model}. 
\begin{table}[h!]
	\centering
	\caption{Parameter values to generate the frequency-dependent reference surface impedances.}
	\begin{tabular}{c c c c c c c}
		\hline
		Parameter & $Z_{1}$ & $Z_{2}$ & $Z_{3}$ & $Z_{4}$ & $Z_{5}$ & $Z_{6}$ \\
		\hline
		$R$      & 0.15 & 0.25 & 1.00 & 0.70 & 1.20 & 0.80 \\
		$K$      & 0.85 & 0.75 & 0.40 & 0.55 & 0.40 & 0.40 \\
		$G$      & 0.015 & 0.020 & 0.100 & 0.100 & 0.200 & 0.400 \\
		$\gamma$ & -0.20 & -0.15 & -0.50 & -0.55 & -0.50 & -0.55 \\
		\hline
	\end{tabular}
	\label{tab:true_parameters}
\end{table}
The parameter values of this reference case, given in Tab.~\ref{tab:true_parameters}, are chosen within the parameter ranges reported by Mondet et al.~\cite{Mondet.2020} Fig.~\ref{fig:Priors+True_Z} presents the reference curves of the real and imaginary parts of the surface impedance in the frequency range from \SI{45}{Hz} to \SI{500}{Hz}, with each impedance surface represented by a different color. A comparison of the curves reveals that $Z_1$ and $Z_2$ consistently display lower absorptive behavior across the entire frequency range, which is reflected in their lower resistance and reactance compared to the other surfaces.
The synthetic measurement data and training datasets are generated with a FE model of the enclosure, implemented in the commercial software COMSOL Multiphysics\textsuperscript{\textregistered} (COMSOL Inc., Stockholm, Sweden). To avoid the problem of an inverse crime, different mesh resolutions are employed. The ground-truth sound pressure field based on the reference surface impedances is computed using tetrahedral elements of quadratic order with a discretization of 30 elements per wavelength at the maximum frequency under study. For training data generation, a coarser discretization of 10 elements per wavelength is applied. To account for measurement and modeling uncertainties in the simulated experiment, zero-mean Gaussian noise is superimposed on the synthetically generated data. The noise is applied to the real and imaginary components of the sound pressure independently, with its level adjusted for each frequency under study according to the specified signal-to-noise ratio (SNR). Guided by the parameter study presented in Sec.~\ref{subsec:Parameter studies}, an SNR of \SI{30}{dB} is adopted throughout this work.

A randomized selection procedure is employed to determine $N_{\text{pos}}$ microphone observation positions while ensuring a prescribed minimum spatial distance between all selected points, enforcing spatial separation. Following this scheme, $N_{\text{pos}}=26$ observation points are selected with a minimum distance of \SI{0.3}{m}, in accordance with the findings of the parameter study presented in Sec.~\ref{subsec:Parameter studies}. These positions are marked as black dots in Fig.~\ref{fig:Model_Cuboid_room}.

The prior distributions of the impedance model parameters are assigned as broad uniform distributions, following the principle of maximum entropy~\cite{Xiang.2020}. 
\begin{table}[h!]
	\centering
	\caption{Assigned prior distributions of the model parameters for the six impedance surfaces.}
	\begin{tabular}{c c c}
		\hline
		Parameter & $Z_{1}$ \& $Z_{2}$ & $Z_{3}$ - $Z_{6}$ \\
		\hline
		$R$      & $\mathcal{U}(0.05,\,0.50)$ & $\mathcal{U}(0.10,\,2.00)$ \\
		$K$      & $\mathcal{U}(0.10,\,1.50)$ & $\mathcal{U}(0.01,\,1.00)$ \\
		$G$      & $\mathcal{U}(0.005,\,0.10)$ & $\mathcal{U}(0.01,\,0.60)$ \\
		$\gamma$ & $\mathcal{U}(-0.50,\,0.00)$ & $\mathcal{U}(-0.70,\,-0.20)$ \\
		\hline
	\end{tabular}
	\label{tab:prior_bounds}
\end{table}
\begin{figure*}[htb]
	\centering
	\includegraphics[width=\textwidth]{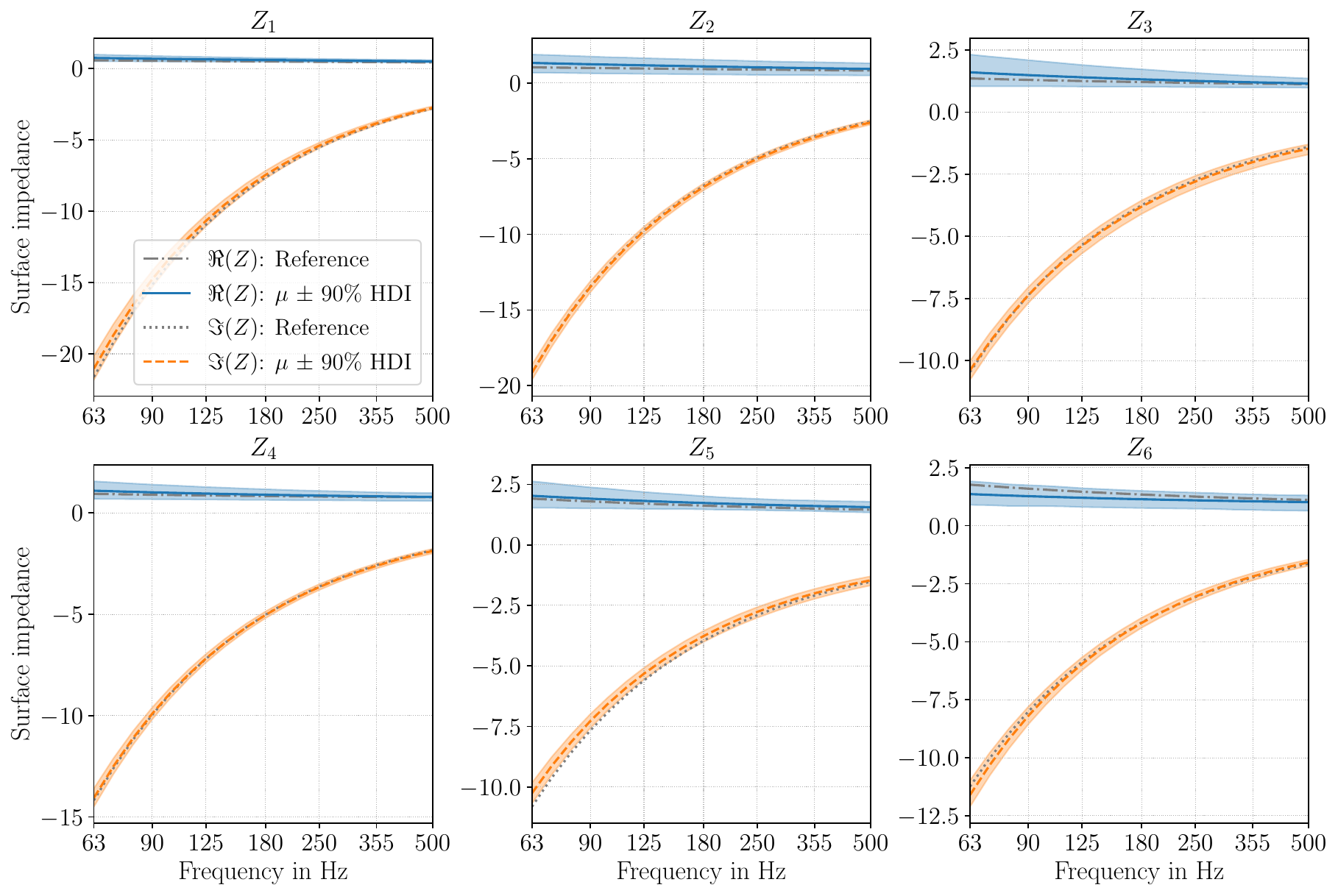}
	\caption{Surface impedance estimates of the proposed SBI framework of $Z_1$--$Z_6$ for the cuboid room. The posterior mean $\mu$ and the 90$\%$ highest density interval (shaded areas) are shown, with the real part depicted in blue, the imaginary part in orange, and the reference values in gray.}
	\label{fig:Cuboid_Impedance_estimate}
\end{figure*}
The corresponding lower and upper bounds, listed in Tab.~\ref{tab:prior_bounds}, are selected to cover a sufficiently wide range while remaining consistent with the parameter ranges reported by Mondet et al.~\cite{Mondet.2020}. Since $Z_{1}$ and $Z_{2}$ exhibit lower absorption than the remaining surfaces, this characteristic is explicitly accounted for in the assigned prior bounds. In practical applications, this information can typically be deduced from the type and choice of the boundary materials, as common construction elements (e.g., glass or concrete) are known to provide substantially lower absorption compared to acoustic treatments such as porous absorbers or perforated panels. Propagating the priors for the model parameters through the impedance model yields the prior distributions of the real and imaginary parts of the surface impedance, with their percentiles shown as shaded regions in Fig.~\ref{fig:Priors+True_Z}. This confirms that the assigned prior bounds span a sufficiently broad range, ensuring comprehensive coverage of plausible impedance values. A prior predictive check is conducted as a diagnostic to evaluate the adequacy of the model and the chosen prior distributions. The results, not shown here for brevity, confirm that the observed data $\boldsymbol{p}_{\text{obs}}$ are well aligned with the distribution of simulated data drawn from the priors and subsequently employed for SBI training.

For training the inference network, $N_{\text{sim}}=4000$ simulations are carried out, in accordance with the parameter study presented in Sec.~\ref{subsec:Parameter studies}. The generation of training data represents the dominant computational cost of the SBI framework. In this work, producing the full training dataset and training the inference network on a office workstation required approximately 10.5 hours, making it feasible to generate all training data within a single overnight run. Once the neural density estimator is trained, posterior sampling becomes highly efficient. Drawing $N_{\text{s}}=100,000$ samples conditioned on the observed sound pressure data requires only a few seconds.

Accurate inference requires a carefully designed network architecture and well-chosen hyperparameters. To identify a suitable configuration, an extensive series of parameter studies and hyperparameter optimizations are conducted. The resulting setup models the neural posterior with masked autoregressive flows~\cite{Papamakarios.2017} combined with rational–quadratic spline coupling layers~\cite{Durkan.2019}, which offer both high expressiveness and numerical stability for approximating complex, high-dimensional distributions. The hyperparameter optimization yields an architecture which consists of $9$ sequential transforms with $11$ spline bins each, where every transform employs fully connected subnetworks of $71$ hidden features. Training is performed with a validation split of $10\%$, a batch size of $200$, $10$ atoms, and a learning rate of $4.2\times10^{-4}$. The SBI framework is implemented in Python using the open-source package \textit{sbi}~\cite{TejeroCantero.2020}.

\begin{figure*}[htb]
	\centering
	\includegraphics[width=\textwidth]{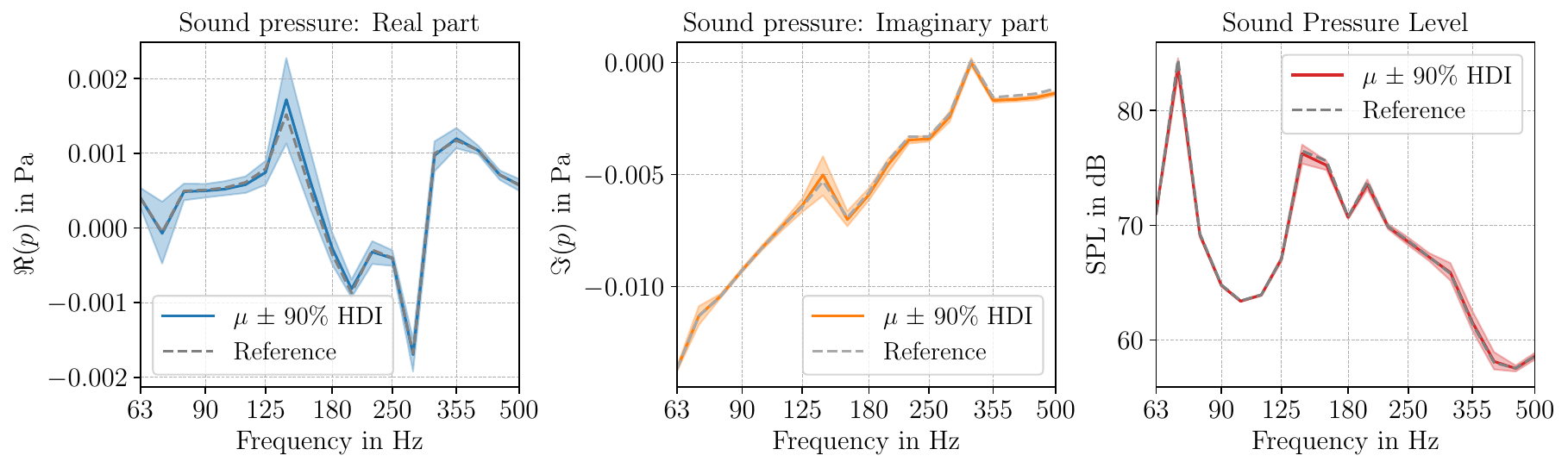}
	\caption{Posterior predictive check results for the cuboid room. The posterior predictive mean $\mu$ (solid lines) and the $90\%$ HDIs (shaded areas) for the real part of the sound pressure $\Re(p)$ (left), the imaginary part $\Im(p)$ (middle), and the sound pressure level (right) are presented. The gray dashed line denotes the synthetic reference sound field averaged over all validation points.}
	\label{fig:Cuboid_PPC_results}
\end{figure*}
For the cuboid room, the surface impedance inference is performed at standardized 1/6-octave band center frequencies ranging from \SI{63}{Hz} to \SI{500}{Hz}. This frequency range encompasses the first 40 room modes, thereby ensuring that the analysis captures the dominant modal behavior of the enclosure. 

The results for all six impedance surfaces $Z_1$--$Z_6$ are presented in Fig.~\ref{fig:Cuboid_Impedance_estimate}. The posterior mean $\mu$ and the $90\%$ highest density interval (HDI) are shown, with the real part in blue, the imaginary part in orange, and the reference values in gray. The HDI is defined as the narrowest interval containing $90\%$ of the posterior probability mass. In contrast to the standard deviation, the HDI provides a more informative uncertainty measure for non-Gaussian distributions, as it directly captures the region of highest probability without assuming symmetry with respect to the mean. The results in Fig.~\ref{fig:Cuboid_Impedance_estimate} demonstrate that the proposed inference framework is able to accurately recover the frequency-dependent reference surface impedances for all six boundaries $Z_1$--$Z_6$. Both, the real and imaginary parts of the impedance are well captured across the entire investigated frequency range, with the posterior mean closely following the reference values. Moreover, the $90\%$ HDIs remain narrow, indicating high confidence in the estimates. The slightly larger deviations and increased uncertainties observed at the lowest frequencies can be explained by the limited information content of the acoustic field in this frequency range. Here, the sound field inside the room is primarily governed by global resonant modes or quasi-uniform pressure distributions, which are only weakly influenced by small variations of the surface impedance. Consequently, sensitivity to impedance changes is low at these frequencies. Before proceeding to the quantitative error analysis in Sec.~\ref{subsec:Error analysis}, the posterior diagnostics introduced in Sec.~\ref{subsec:Posterior diagnostics} are employed to assess the quality and calibration of the trained posterior estimator.

\subsection{\label{subsec: Posterior validation} Posterior validation}
As described in Sec.~\ref{subsec:Posterior diagnostics}, the quality and reliability of the posterior estimator are evaluated by means of a PPC. Specifically, $N_{\text{ppc}}=1000$ samples are drawn from the inferred posterior, passed through the impedance model and the forward simulator, and evaluated at a validation set of 2361 mesh nodes that have not been used during training. The results are summarized in Fig.~\ref{fig:Cuboid_PPC_results}, which shows the posterior predictive mean together with the $90\%$ HDI across the considered frequency range for the real part $\Re(p)$ (left), the imaginary part of the sound pressure $\Im(p)$ (middle), and the sound pressure level (SPL) in \si{dB} relative to $p_0 = \SI{2e-5}{Pa}$ (right). The synthetic reference sound field, averaged over all validation points, is indicated by the gray curve. Across the examined frequency range, the posterior predictive mean exhibits close agreement with the reference, and the reference values consistently lie within the $90\%$ HDIs. This shows that the inferred impedance posterior, in combination with the simulator model, accurately captures both the amplitude and phase characteristics of the sound field. The predictive intervals broaden in the vicinity of pronounced modal frequencies (e.g., around \SI{71}{Hz} and \SI{140}{Hz}), highlighting the increased sensitivity of the acoustic response to boundary conditions in these frequency regions, while remaining narrow in less resonant bands. A quantitative evaluation of the predictive accuracy is provided in Sec.~\ref{subsec:Error analysis} based on the modal assurance criterion.

\begin{figure}[htb]
	\centering
	\includegraphics[width=0.95\columnwidth]{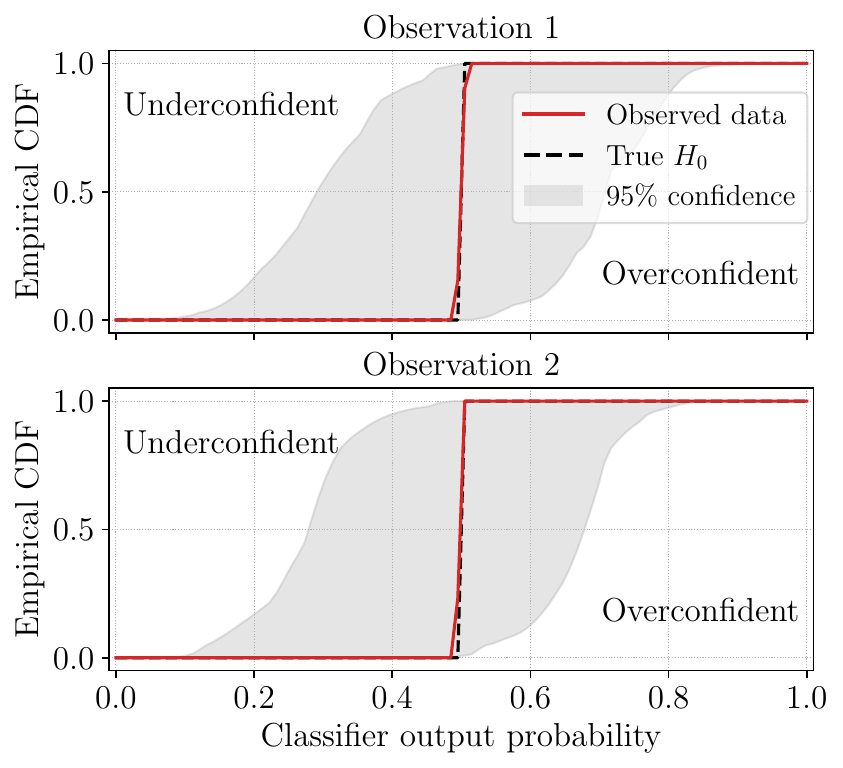}
	\caption{Results of the L-C2ST coverage diagnostic for two randomly chosen observation samples from the validation dataset. The empirical CDFs of classifier output probabilities (red) compared to the expected distribution under perfect calibration (black dashed) are shown, with the 95\% confidence region indicated in gray.}
	\label{fig:Cuboid_LC2ST_Plot}
\end{figure}

Fig.~\ref{fig:Cuboid_LC2ST_Plot} shows the results of the L-C2ST coverage diagnostic introduced in Sec.~\ref{subsec:Posterior diagnostics}. 
The plots display the empirical cumulative distribution functions (CDFs) of classifier output probabilities for two randomly picked observation samples from the validation dataset (red lines), compared against the expected distribution under the null hypothesis $H_0$ of perfect calibration (black dashed lines). The gray shaded regions indicate the 95\% confidence bounds, within which classifier outputs are statistically indistinguishable from a uniform distribution and thus correspond to a well-calibrated posterior. For both observation samples, the empirical CDFs closely track the theoretical step function at 0.5 and remain largely within the confidence intervals, demonstrating that the estimated posterior distributions are statistically well-calibrated and show no systematic tendency toward under- or overconfidence. Overall, both the PPC and the L-C2ST diagnostic confirm that the proposed inference procedure yields posterior estimates that are statistically well-calibrated and physically consistent.  

\subsection{\label{subsec:Impedance tube measurements} Impedance tube measurements}
For the results presented above, the reference impedance curves have been generated using a predefined set of impedance model parameters, which is given in Tab.~\ref{tab:true_parameters}. While this approach is valuable for verifying the inference framework, it represents a simplification, as the impedance model is assumed to be perfectly satisfied up to the added noise level. To address this limitation and account for uncertainties arising from the impedance model, experimentally measured surface impedances are employed as reference data. These measurements are obtained using a standardized two-microphone impedance tube setup following the transfer function method~\cite{ISO105342.1998}. 
\begin{figure}[htb]
	\centering
	\includegraphics[width=\columnwidth]{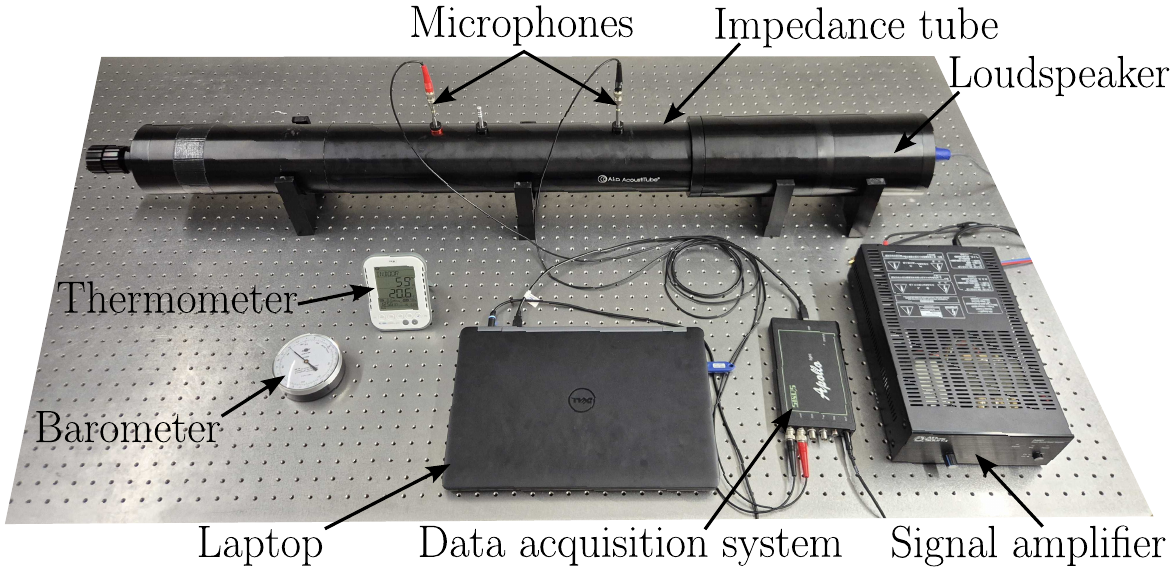}
	\caption{Two-microphone impedance tube measurement setup following the ISO~10534-2 standard.}  
	\label{fig:Impedance_tube}
\end{figure}
The material under investigation is a Flexi A50 acoustic absorber (Vicoustics, Paços de Ferreira, Portugal), which is commercially used in recording studios and practice booths. The complete experimental setup is illustrated in Fig.~\ref{fig:Impedance_tube}. The measurements are carried out using an AED AcoustiTube system (Gesellschaft für Akustikforschung Dresden mbH, Dresden, Germany) with an inner diameter of \SI{100}{mm} and an effective frequency range of \SI{58}{Hz} to \SI{2}{kHz}. The setup uses two 1/4” microphones of type M370 (Microtech Gefell GmbH, Gefell, Germany), which are calibrated prior to the experiments using a pistonphone type 4228 (Brüel \& Kjær, Virum, Denmark). Data acquisition is performed with an Apollo Light system (SINUS Messtechnik GmbH, Leipzig, Germany), and the excitation signal is driven by a single-channel power amplifier PA601 (AtlasIED, Phoenix, USA). During the measurements, environmental conditions have been monitored and remained stable at a temperature of \SI{20.6}{\celsius}, relative humidity of \SI{59}{\%}, and a barometric pressure of \SI{967}{hPa}. 
\begin{figure}[htb]
	\centering
	\includegraphics[width=0.7\columnwidth]{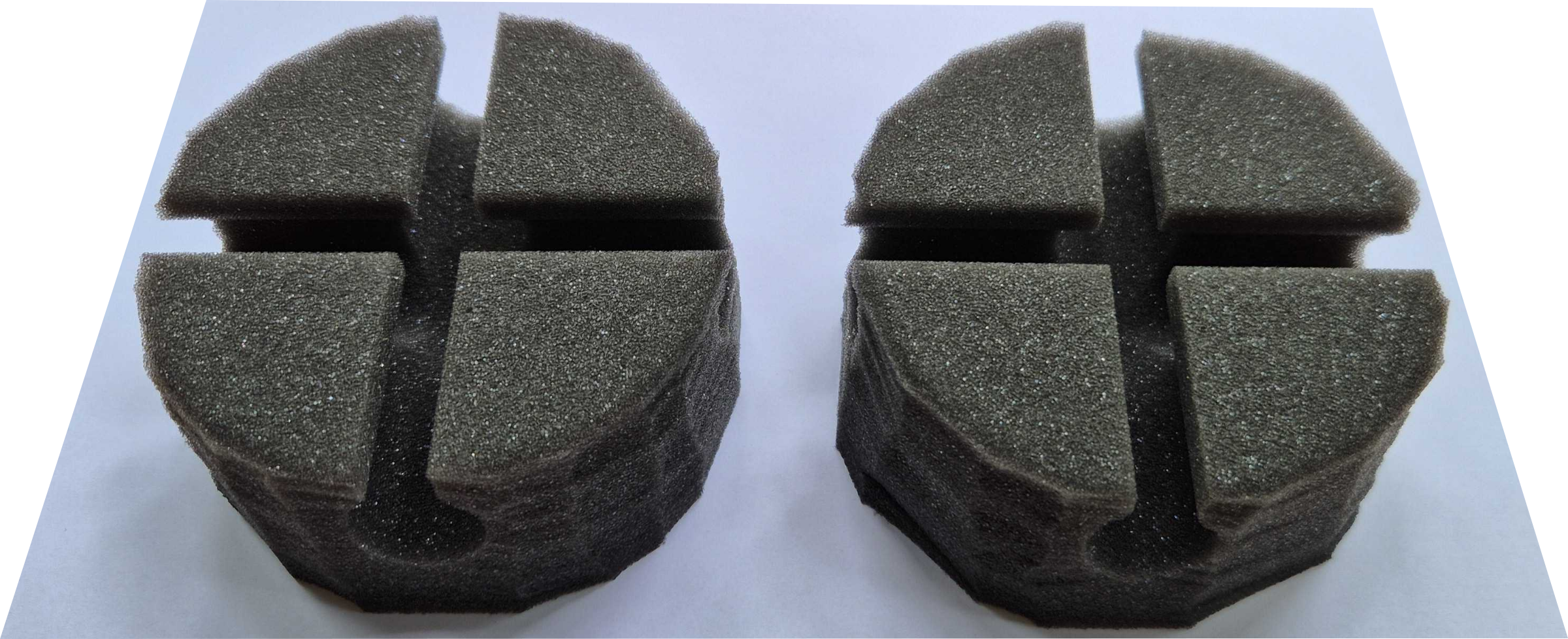}
	\caption{Two samples of the Vicoustics Flexi A50 absorption material prepared for testing in the impedance tube.}  
	\label{fig:A50_Samples}
\end{figure}
The two test samples, shown in Fig.~\ref{fig:A50_Samples}, are cut to a diameter of \SI{100}{mm} to ensure a precise fit within the impedance tube. To account for possible variability due to mounting conditions, each of the two samples have been measured three times with repeated dismounting and remounting. The resulting surface impedance curves, presented as gray solid lines in Fig.~\ref{fig:Priors+True_Z}, correspond to the mean of these repeated measurements and thus provide a robust and reliable reference for the subsequent analysis. The experimentally measured surface impedances are assigned as reference values for three wall surfaces, namely $Z_{3}^\ast$, $Z_{5}^\ast$, and $Z_{6}^\ast$. To prevent overly homogeneous boundary conditions, which would not reflect realistic acoustic environments, the remaining walls $Z_1$, $Z_2$, and $Z_4$ are kept identical to the synthetic configuration described earlier. This mixed setup combines measured and synthetic impedances, ensuring both realism and sufficient variability across the room boundaries.  

\subsection{\label{subsec:Error analysis} Error analysis}
To quantitatively assess the accuracy of the inferred surface impedances relative to both synthetic and measured references, the relative $L_2$ norm error of the complex impedance is defined as  
\begin{align}
	\epsilon_{2} \;=\; \frac{1}{N_f} 
	\sum_{f=1}^{N_f}
	\sqrt{ \frac{ \big|  \boldsymbol{Z}_{\text{SBI}}(f) -  \boldsymbol{Z}_{\text{ref}}(f) \big|^2 }
		{ \big|  \boldsymbol{Z}_{\text{ref}}(f) \big|^2 } },
	\label{eq:L2_error}
\end{align}
with $N_f$ denoting the number of investigated frequencies. This error measure captures the normalized deviation between the inferred frequency-dependent, complex-valued surface impedances $ \boldsymbol{Z}_{\text{SBI}}(f)$ and the reference $ \boldsymbol{Z}_{\text{ref}}(f)$, averaged across frequency.  
\begin{figure}[htb]
	\centering
	\includegraphics[width=0.95\columnwidth]{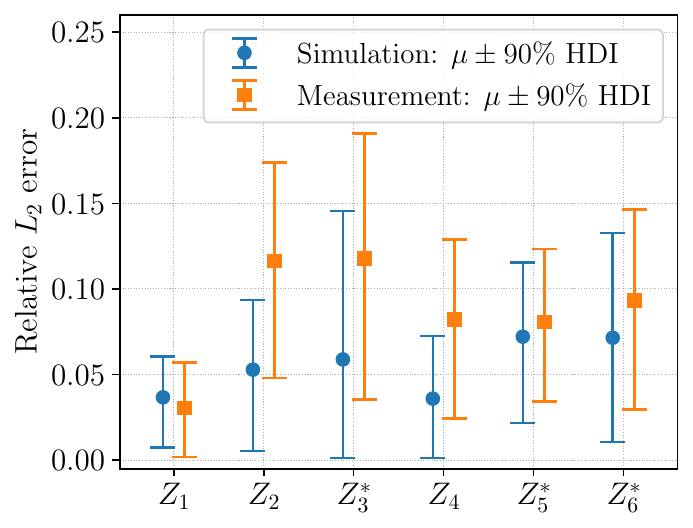}
	\caption{Relative $L_2$ norm errors for the surface impedances $Z_1$, $Z_2$, and $Z_4$ using synthetic (blue) and $Z_{3}^\ast$, $Z_{5}^\ast$, and $Z_{6}^\ast$ using measured (orange) reference surface impedance data, shown as mean values with $90\%$ HDIs.}  
	\label{fig:Cuboid_L2_Z_errorbar}
\end{figure}
Fig.~\ref{fig:Cuboid_L2_Z_errorbar} shows the mean values and the corresponding $90\%$ HDIs over all posterior samples of the relative $L_2$ norm error for the surface impedances $Z_1$, $Z_2$, and $Z_4$ using synthetic (blue) and $Z_{3}^\ast$, $Z_{5}^\ast$, and $Z_{6}^\ast$ using measured (orange) reference surface impedance data. For the synthetic references, the errors are consistently low, with mean values below 0.08 across all cases. When real measured impedances are used as reference, the errors are slightly higher on average and display broader uncertainty ranges. This increase reflects both the deviations of the real material behavior from the impedance model and the additional experimental uncertainties inherent in the measurement process. Nevertheless, the errors remain moderate across all cases with mean $L_2$ errors below $0.13$, demonstrating that the proposed inference framework generalizes robustly from idealized to measured data and is applicable to realistic acoustic materials beyond synthetic test cases.

To evaluate the accuracy of the posterior predictive behavior obtained from the PPC, the modal assurance criterion (MAC) is employed. It is defined as  
\begin{align}
	\mathrm{MAC}\;=\; 
	\frac{\left| \boldsymbol{p}_{\text{ref}}^{H} \, \boldsymbol{p}_{\text{SBI}} \right|^{2}}
	{\left( \boldsymbol{p}_{\text{ref}}^{H} \boldsymbol{p}_{\text{ref}} \right) \left( \boldsymbol{p}_{\text{SBI}}^{H} \boldsymbol{p}_{\text{SBI}} \right)},
	\label{eq:MAC}
\end{align}
where $\boldsymbol{p}_{\text{ref}}$ denotes the complex-valued reference sound pressure vector, $\boldsymbol{p}_{\text{SBI}}$ the PPC prediction evaluated at the validation positions. The superscript $H$ indicates the Hermitian transpose. 
The MAC quantifies the degree of correlation between two spatial fields and is a well-established metric for assessing spatial similarity. It simultaneously accounts for amplitude and phase, making it particularly suitable for comparing complex-valued acoustic pressure fields. The MAC takes values between 0 and 1, with 0 indicating no similarity and 1 denoting perfect agreement. In the present context, it provides a rigorous measure of how well the PPC-predicted sound pressure fields reproduce the spatial structure of the reference sound field.  
\begin{figure}[htb]
	\centering
	\includegraphics[width=0.91\columnwidth]{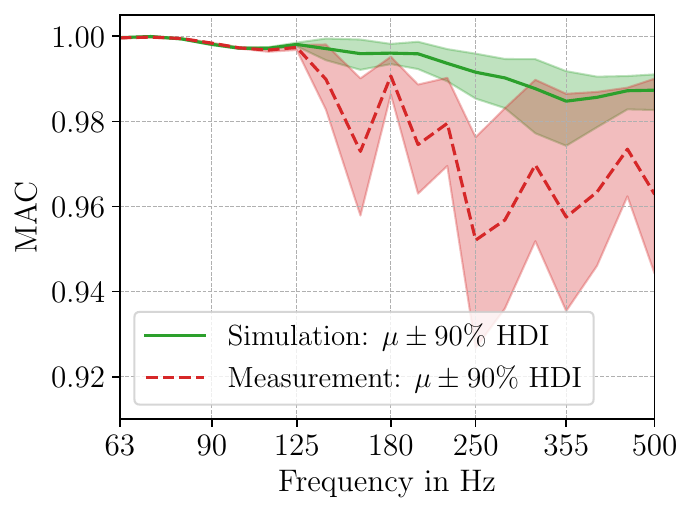}
	\caption{MAC of the posterior predictive sound pressure fields using synthetic (green) and measured (red) reference surface impedances, shown as mean values with $90\%$ HDIs.}  
	\label{fig:Cuboid_PPC_MAC}
\end{figure}
Fig.~\ref{fig:Cuboid_PPC_MAC} presents the mean and the $90\%$ HDIs of the MAC, evaluated over all PPC samples as a function of frequency, for both synthetic (green) and measured (red) reference impedances. In the synthetic case, the MAC remains nearly equal to unity across the full frequency range with very narrow uncertainty bounds, demonstrating an almost perfect reproduction of the spatial sound pressure fields. 
\begin{figure*}[htb]
	\centering
	\includegraphics[width=\textwidth]{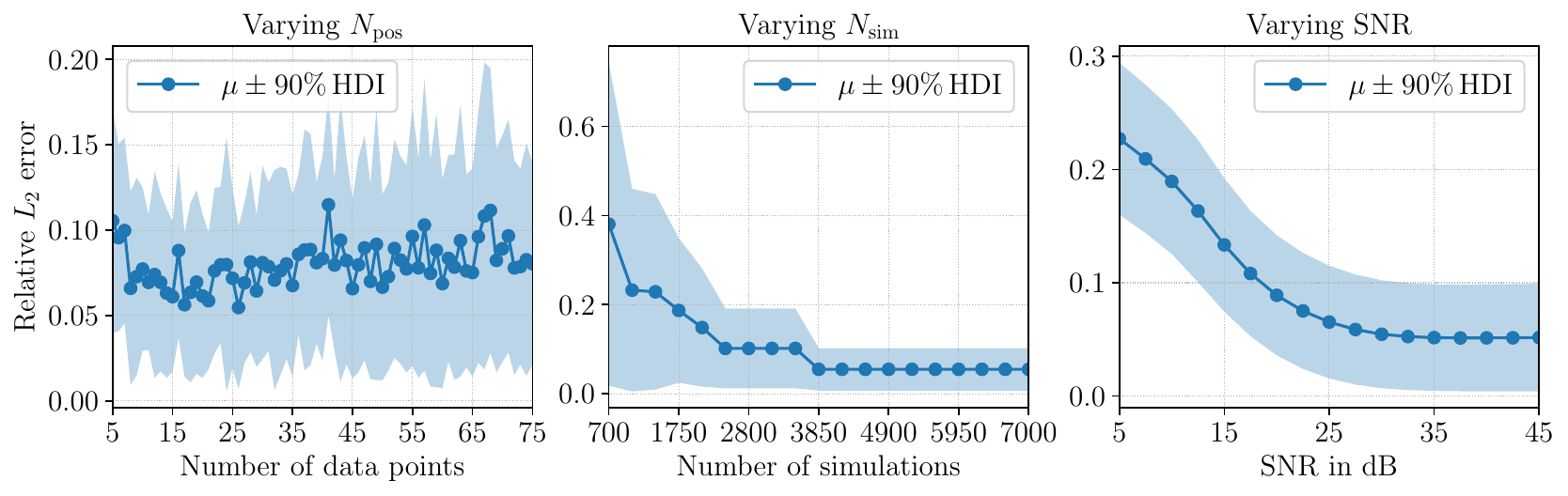}
	\caption{Parameter study showing the influence of the number of observation points $N_{\text{pos}}$ (left), the number of training simulations $N_{\text{sim}}$ (middle), and the SNR (right) on the mean relative $L_2$ error and $90\%$ HDI.}
	\label{fig:Cuboid_HPO_results}
\end{figure*}
Using measured impedances as reference, the MAC values remain high overall but show a slight reduction, particularly at frequencies above \SI{125}{Hz}. The wider uncertainty intervals in this case reflect the combined effects of measurement noise and deviations of the real material behavior from the idealized impedance model. Despite these differences, the MAC mean values remain consistently above 0.95, confirming that the inferred posterior achieves a high degree of spatial agreement with the reference fields. In particular, at very low frequencies below \SI{125}{Hz}, the agreement is nearly perfect, with only minimal predictive uncertainty.

\subsection{\label{subsec:Parameter studies} Parameter studies}
To evaluate the sensitivity of the proposed method to variations in key parameters of the framework, a series of parameter studies has been conducted. 
Fig.~\ref{fig:Cuboid_HPO_results} presents the effect of these parameters on the mean relative $L_2$ error and $90\%$ HDIs. The analysis of the number of observation points $N_{\text{pos}}$ (left) indicates an optimal configuration at approximately 26 points, beyond which additional observations no longer improve accuracy, reflecting a saturation in the information gained from further measurement data. 
In contrast, increasing the number of training simulations $N_{\text{sim}}$ (middle) substantially reduces both error and uncertainty, with the most significant improvements occurring up to about 3850 simulations. Beyond this threshold, the results consistently converge toward an identical solution, confirming that the framework has effectively reached convergence. Finally, the robustness analysis with respect to the SNR of the added Gaussian noise (right) shows that higher SNR values systematically enhance accuracy, while the level of uncertainty remains essentially unchanged. Notably, the performance stabilizes already at moderate SNR values around \SI{25}{dB}, indicating that the method remains reliable even under comparatively high noise conditions. This ability of the framework to maintain predictive accuracy under noisy conditions highlights its suitability for real-world acoustic scenarios.

\section{\label{sec:Car cabin problem}Car cabin problem}
\begin{figure}[tbh]
	\centering
	\includegraphics[width=\columnwidth]{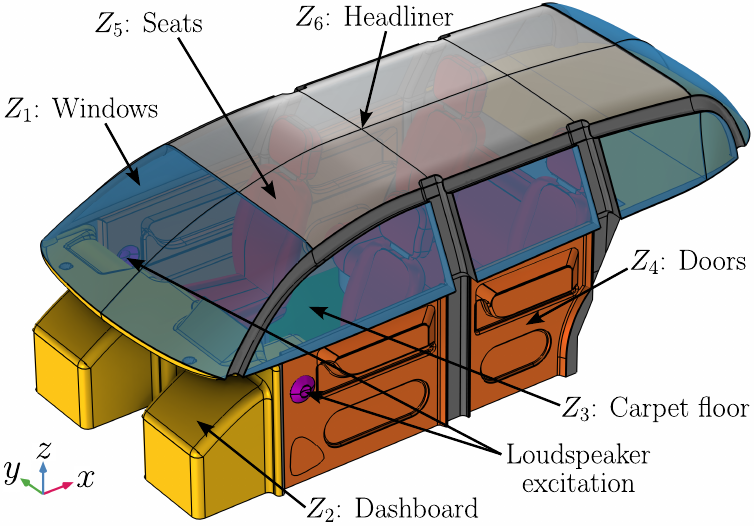}
	\caption{Geometric model of the car cabin with surface impedance regions highlighted in different colors. Dark gray areas denote sound-hard surfaces, while the loudspeaker membrane used for excitation is shown in pink.}
	\label{fig:Car_cabin_model}
\end{figure}

Building on the benchmark cuboid room example, the proposed method is subsequently applied to a considerably more complex three-dimensional car cabin model with realistic geometric features.  
This model corresponds to an example case study provided in COMSOL Multiphysics\textsuperscript{\textregistered}. The dimensions of the cabin are $L_x=\SI{3.44}{m}$, $L_y=\SI{1.60}{m}$, and $L_z=\SI{1.22}{m}$, corresponding to a total volume of $\SI{3.36}{m^3}$. The model is illustrated in Fig.~\ref{fig:Car_cabin_model}, where the surfaces with different impedance conditions are highlighted in distinct colors: $Z_1$ for the windows (light blue), $Z_2$ for the dashboard (yellow), $Z_3$ for the carpet floor (green), $Z_4$ for the doors (orange), $Z_5$ for the seats (purple), and $Z_6$ for the headliner (light gray). The reference surface impedance values are identical to those specified in Tab.~\ref{tab:true_parameters}, while the dark gray regions in Fig.~\ref{fig:Car_cabin_model} are modeled as perfectly reflecting (sound hard) surfaces. The sound field excitation is introduced via a prescribed normal velocity of $v_n = \SI{0.01}{\meter \per \second}$ at the loudspeaker membrane, shown in pink in Fig.~\ref{fig:Car_cabin_model}. All remaining model parameters are consistent with those described in Sec.~\ref{sec: Cuboid room}.
Parameter studies, not reported here for brevity, indicate that an optimal configuration is achieved with $N_{\text{pos}}=28$ observation points, subject to a minimum spacing of \SI{0.24}{\meter} between positions. In addition, the studies show that the inference framework exhibits stable convergence with $N_{\text{sim}}=4000$ training simulations and a signal-to-noise ratio (SNR) of \SI{30}{dB}. The prior distributions for the impedance model parameters are identical to those specified in Tab.~\ref{tab:prior_bounds} and the network architecture and hyperparameter settings of the inference model are consistent with those described in Sec.~\ref{sec: Cuboid room}. The inference is performed at standardized 1/6-octave center frequencies between \SI{45}{Hz} and \SI{400}{Hz}, covering more than the first 50 cavity modes and thereby ensuring that the analysis captures the dominant modal behavior of the enclosure. For the car cabin example, generating the full training dataset and training the inference network requires approximately 38 hours on an office workstation.
\begin{figure}[htb]
	\centering
	\includegraphics[width=0.92\columnwidth]{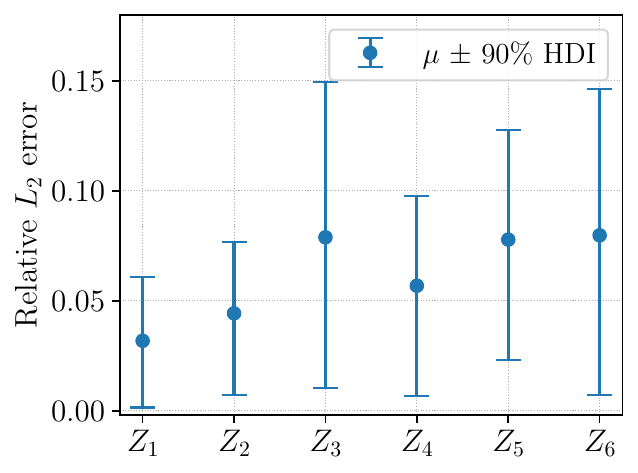}
	\caption{Relative $L_2$ error with $90\%$ HDIs for the six estimated surface impedances $Z_1$--$Z_6$ of the car cabin model.}
	\label{fig:Car_cabin_L2_Z_errorbar}
\end{figure}

Fig.~\ref{fig:Car_cabin_L2_Z_errorbar} presents the mean relative $L_2$ error together with the corresponding $90\%$ HDIs for the six surface impedances $Z_1$--$Z_6$. 
Across all surfaces, the errors remain below 0.08, demonstrating a level of accuracy comparable to the cuboid room case. The largest error and widest uncertainty interval are observed for $Z_3$ (carpet floor), which can be attributed to its comparatively small surface area and thus reduced influence on the overall acoustic field.
As in the cuboid room study, the quality and reliability of the posterior estimator are assessed using L-C2ST coverage diagnostics and a PPC. The validation set comprises the remaining 29972 mesh nodes that have not been employed as observation points for the inference. Also for the more complex car cabin scenario, the L-C2ST results likewise confirm a well-calibrated posterior estimate without any indication of under- or overconfidence (results not shown here for brevity). The accuracy of the posterior predictive behavior obtained from the PPC is evaluated using the MAC defined in Eq.~\eqref{eq:MAC}.
\begin{figure}[htb]
	\centering
	\includegraphics[width=0.91\columnwidth]{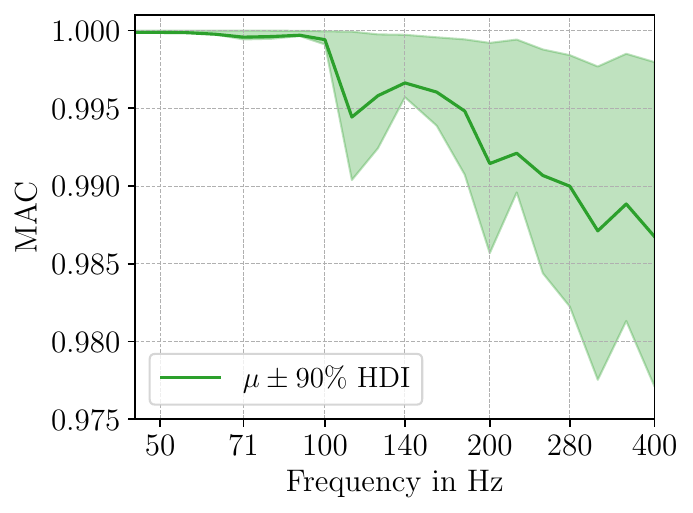}
	\caption{MAC with mean $\mu$ and $90\%$ HDIs for the posterior predictive sound pressure fields of the car cabin model across the frequency range.}
	\label{fig:Car_cabin_PPC_MAC}
\end{figure}
Fig.~\ref{fig:Car_cabin_PPC_MAC} shows the mean and the $90\%$ HDI of the MAC, evaluated over all PPC samples. Across the considered frequency range, the MAC remains consistently high with values above 0.985, indicating excellent agreement between the predicted and reference sound pressure fields. 
A slight degradation in MAC can be observed at higher frequencies, accompanied by wider uncertainty intervals, which reflects the increased sensitivity of the model to variations in boundary conditions and measurement noise. Nonetheless, the overall results demonstrate that the inferred surface impedance posterior distributions achieve a robust reproduction of the acoustic field inside the car cabin.

\section{\label{sec:Conclusions} Conclusions and Outlook}
This study has introduced a Bayesian framework for the \textit{in situ} estimation of frequency-dependent, piece-wise constant acoustic surface impedances in small enclosures, employing simulation-based inference (SBI). Bayesian inference is a well-established method that naturally incorporates prior knowledge and enables rigorous uncertainty quantification. However, conventional sampling-based Bayesian approaches become impractical for inverse problems that are high-dimensional or involve computationally demanding forward models. SBI specifically addresses these challenges of high dimensional parameter inference, by leveraging the expressiveness of modern neural network architectures. The inference network learns a direct mapping from simulated data to the posterior distributions of the unknown parameters, thereby eliminating the need for an explicit likelihood formulation. Once trained, the network can efficiently generate posterior samples for measured data without requiring additional simulations or expensive sampling schemes. 

To represent the frequency-dependent behavior of the acoustic surface impedance, a generalized damped oscillator model is employed, extended by a fractional calculus term to capture the complex frequency-dependent characteristics more accurately. The posterior distributions of the piece-wise constant surface impedances at different boundaries are obtained by inferring the parameters of this impedance model from a limited set of sound pressure measurements inside the enclosure. 
The proposed method has successfully been applied to a three-dimensional cuboid room model, representative of compact acoustic enclosures such as soundproof phone cabins commonly used in office environments or recording booths in music studios. The inferred impedance values show excellent agreement with the reference data across the considered frequency range from \SI{63}{Hz} to \SI{500}{Hz}, accurately capturing all six individual surface impedances with an average maximum relative $L_2$ norm error of 0.08. To further validate the approach, impedance tube measurement data are employed as reference impedances to generate synthetic measurements inside the cuboid room, thereby relaxing the assumption that the impedance model is perfectly satisfied. Also in this case, the framework achieves high accuracy, demonstrating its robustness and ability to generalize from idealized simulations to real measurement data. 

Moreover, the method has been applied to a more complex three-dimensional car cabin model in the frequency range from \SI{45}{Hz} to \SI{400}{Hz}. Even for this considerably more challenging configuration, the approach achieves accurate surface impedance estimation and reliable uncertainty quantification, with an average maximum relative $L_2$ norm error of 0.08. 
To assess the quality and reliability of the posterior estimator, a posterior predictive check together with the local classifier two-sample test has been employed as coverage diagnostics. The results confirm accurate posterior predictive performance and statistically well-calibrated posterior distributions, showing no systematic tendency toward under- or overconfidence in either of the investigated cases.

Nevertheless, SBI also presents certain challenges, as it relies on the availability of a physically validated simulation model to generate training data. Consequently, it is particularly advantageous in contexts where detailed models are readily available, such as digital twin frameworks or model updating applications. While the data generation process for complex simulations can be computationally expensive, it only needs to be carried out once in an offline stage, after which the trained inference model can be applied to new measurements without incurring additional simulation costs.

Future work will focus on extending the proposed framework to more complex acoustic models, including geometries with higher detail and more heterogeneous boundary conditions. An important direction is the inference of spatially varying surface impedances, which allow for a more accurate representation of materials with non-uniform acoustic properties. Finally, experimental validation through dedicated \textit{in situ} measurements in small enclosures is planned, providing a crucial step toward demonstrating the practical applicability and robustness of the framework under real measurement conditions.

\begin{acknowledgments}
The authors would like to thank their colleagues from the material characterization research group for the valuable discussions. 
\end{acknowledgments}

\section*{\label{sec:Literature} Literature}
\bibliography{bibliography}

\end{document}